\documentclass[a4paper,10pt,twoside]{cpc-hepnp}

\usepackage{subfigure}
\usepackage{color}
\usepackage{epsfig}
\usepackage{indentfirst}
\usepackage{multirow}
\usepackage{bm}
\usepackage{fancyhdr}
\usepackage{enumerate}
\usepackage[perpage,symbol]{footmisc}
\usepackage[bookmarksnumbered,pdfstartview=FitH,colorlinks=true,menucolor=black,linkcolor=blue]{hyperref}
\usepackage{multicol}
\usepackage{booktabs}
\usepackage{amssymb,bm,mathrsfs,bbm,amscd}
\usepackage[tbtags]{amsmath}
\usepackage{lastpage}
\usepackage{CJK}

\begin{document}

\fancyhead[c]{\small Chinese Physics C~~~Vol. X, No. X (X) X}
\fancyfoot[C]{\small X-\thepage}

\footnotetext[0]{Received X X X}

\title{Combining upper limits with a Bayesian approach\thanks{This article
    is supported in part by National Natural Science Foundation of
    China (NSFC) under Contract No. 11005115, the Joint Funds of the
    NSFC and Chinese Academia Sinica's Big Scientific Apparatus under
    Contract No. U1232109.}}

\author{%
 Liu Yang$^{1,2}$%
\quad Kai Zhu$^{2}$\email{zhuk@ihep.ac.cn}%
\quad Yongsheng Zhu$^{2}$%
\quad Hao Cai$^{1}$\email{hcai@whu.edu.cn}%
}
\maketitle

\address{%
$^1$ Department of Physics, School of Physical Science and
Technology, Wuhan University, 430072 Wuhan,
People's Republic of China \\
$^2$ Institute of High Energy Physics, Chinese Academy of Science,
PO Box 918(1), Beijing 100049, People's Republic of China \\
}

\begin{abstract}
We discuss how to determine and combine upper limits based on observed events and
estimated backgrounds with a Bayesian method, when insignificant signals are observed in
independent measurements.  In addition to some general features deduced from the
analytical formulae, systematic numerical results are obtained by a C$++$ program (CULBA)
for low-count experiments, which can be used as a reference to combine two
upper limits.
\end{abstract}

\begin{keyword}
upper limit, Bayesian method, combine results
\end{keyword}

\begin{pacs}
02.50.Cw, 02.50.Tt, 02.70.Pt
\end{pacs}

\footnotetext[0]{\hspace*{-3mm}\raisebox{0.3ex}{$\scriptstyle\copyright$}2013
Chinese Physical Society and the Institute of High Energy Physics of
the Chinese Academy of Sciences and the Institute
of Modern Physics of the Chinese Academy of Sciences and IOP Publishing Ltd}%

\begin{multicols}{2}

\section{Introduction}
How to combine independent results to extract the most
information appropriately is a crucial problem for experimentalists.
Some statistical methods have been proposed~\cite{cern:2000}, and the
Bayesian method is a promising one~\cite{D'Agostini:2003qr}.
Its basic idea is simple; the probability density function (PDF),
$f(\mu|x)$, of a truth parameter $\mu$ deduced from an observational parameter $x$ is read as~\cite{Bayes:1764vd}:
\begin{equation}
\label{eq:bay}
 f(\mu | x) 
= \frac{f(x|\mu)f_0(\mu)}{\int f(x|\mu)f_0(\mu)\mathrm{d}\mu} \;
\end{equation}
in which $f_0(\mu)$, named the prior, is the degree of belief
attributed to $\mu$ before observation, and $f(\mu|x)$, called
 the posterior, corresponding to the prior, is the updated likelihood
that $\mu$ will produce the observed effect $x$.

The Bayesian method has the advantage of combining results since it provide a natural means to
include additional knowledge by adding nuisance parameters, compared with other statistics
methods~\cite{Agashe:2014kda}. However, an obvious weakness of this method is that its
posterior depends on the choice of the prior. Even for a uninformative prior, there are
different proposals such as uniform prior, Haldane prior~\cite{Haldane:1932}, Jefferys
prior ~\cite{Jefferys:1946}, reference prior~\cite{Bernardo:1979} and so
on~\cite{Jaynes:1968}. How to select an appropriate prior is a kind of art. This
situation is even worse for rare processes and small signals, although that
has been discussed somewhat in depth and methods based on the spirit of Bayesian have been
developed~\cite{Narsky:1999kt,Escoubes:1987ft,Zhu:2007zza}. Here experimentalists face
a double risk of missing a real signal or ruining physics sensitivity. In this paper,
we show that, analytically and numerically, by using the first experiment result as the prior
for the second one to combine two upper limits will improve this situation significantly. We
then partially solve this problem.

\section{Combining two probability density functions}
\label{sec:com}
As a starting point, let us consider a counting measurement on the
number of a small signal. In the signal region, x events are
observed, that is a Poisson random variable with average value
$\lambda_S+\lambda_B$, where $\lambda_S$ and $\lambda_B$ are the
expected numbers of signal and background events respectively.
$\lambda_S$ is the signal parameter that one wants to infer, while
$\lambda_B$ is a nuisance parameter, which could be estimated by
background regions or theoretical predictions. From the spirit of the
Bayesian method, it is natural to deduce the probability of
$\lambda_S$ signal~\cite{D'Agostini:2003nk}
\end{multicols}

\begin{equation}
\label{eq:bay2}
  f(\lambda_S | x, f_0(\lambda_S,\lambda_B)) =
  \frac{\int e^{-(\lambda_S+\lambda_B)}(\lambda_S+\lambda_B)^x f_0(\lambda_S,\lambda_B)
    \mathrm{d}\lambda_B}
       {\int \int e^{-(\lambda_S+\lambda_B)}(\lambda_S+\lambda_B)^x f_0(\lambda_S,\lambda_B)
     \mathrm{d}\lambda_S \mathrm{d}\lambda_B}
\end{equation}

Usually the priors of signal and background are independent, {\it
i.e.} $f_0(\lambda_S,\lambda_B) = f_0(\lambda_S)f_0(\lambda_B)$.
Suppose there are two experiments examining the same physics signal
$\lambda_S$, with observation $x_1$ and $x_2$ respectively.  As
pointed out by D'Agostini~\cite{D'Agostini:2003nk} (section 6.3), by
applying Bayes' theorem a second time, {\it i.e.} using the posterior
of the first experiment as the prior for the second one, we obtain
the final posterior PDF for $\lambda_S$, by which the final result
for the $\lambda_S$ of two experiments can be inferred.  However, in
general $\lambda_S$ is related to a parameter ${\cal B}$, which is
the quantity to be measured, by an experimental factor $h$ via
$\lambda_S = h {\cal B}$. For example, in $e^+ e^-$ collision
experiments, if we want to measure a branching ratio $\cal{B}$ for a
decay $R \to f$, then $h = \cal{L}\epsilon$, where $\cal{L}$ and
$\epsilon$ are the accumulated luminosity and the detection
efficiency for $R \to f$ signal events, respectively. Defining $h_1$
and $h_2$ are the experimental factors of the two experiments
respectively, and replacing $f_0(\lambda_S)$ in Eq.~(\ref{eq:bay2})
with $f(\lambda_{S} | x_1, f_0(\lambda_{S},\lambda_{B1}))$, after
some derivation we get
\begin{equation}
\label{eq:ori_main}
  f({\cal B} | x_2, f_0(\lambda_{S2},\lambda_{B_2}))
=
  \frac{\int \int e^{-[(h_1+h_2){\cal B}+\lambda_{B_1}+\lambda_{B_2}]}
  (h_1{\cal B}+\lambda_{B_1})^{x_1} (h_2{\cal B}+\lambda_{B_2})^{x_2} f_0(\lambda_{S1})
  f_0(\lambda_{B_1}) f_0(\lambda_{B_2}) \mathrm{d}\lambda_{B_1} \mathrm{d}\lambda_{B_2}}
       {\int \int \int e^{-[(h_1+h_2){\cal B}+\lambda_{B_1}+\lambda_{B_2}]}
     (h_1{\cal B}+\lambda_{B_1})^{x_1} (h_2{\cal B}+\lambda_{B_2})^{x_2} f_0(\lambda_{S1})
     f_0(\lambda_{B_1}) f_0(\lambda_{B_2}) \mathrm{d}\lambda_{S1}
     \mathrm{d}\lambda_{B_1}  \mathrm{d}\lambda_{B_2}}.
\end{equation}
Here $f_0(\lambda_{S1})$ is only the prior of the first experiment, and it can be set to a
uniform shape $f_0(\lambda_{S1})=k$ ($k$ is a constant) to indicate there is totally no
knowledge before this measurement. For simplicity we normalize this formula to the number
of the signal of the second experiment (renamed as $\lambda_S$):
\begin{equation}
\label{eq:main}
  f(\lambda_S | x_2, f_0(\lambda_S,\lambda_{B_2}))
=
  \frac{\int \int e^{-[(g+1)\lambda_S+\lambda_{B_1}+\lambda_{B_2}]}
  (g\lambda_S+\lambda_{B_1})^{x_1} (\lambda_S+\lambda_{B_2})^{x_2}
  f_0(\lambda_{B_1}) f_0(\lambda_{B_2}) \mathrm{d}\lambda_{B_1} \mathrm{d}\lambda_{B_2}}
       {\int \int \int e^{-[(g+1)\lambda_S+\lambda_{B_1}+\lambda_{B_2}]}
     (g\lambda_S+\lambda_{B_1})^{x_1} (\lambda_S+\lambda_{B_2})^{x_2}
     f_0(\lambda_{B_1}) f_0(\lambda_{B_2}) \mathrm{d}\lambda_{S}
     \mathrm{d}\lambda_{B_1}  \mathrm{d}\lambda_{B_2}},
\end{equation}
where $g \equiv h_1/h_2$ is the normalization factor, that represents the ratio between the experimental
factors of the two experiments. $g \ll 1$ or $g \gg 1$ means one experiment is much more
sensitive than the other one. Due to our study, in this kind of situation, the
final result will be dominated only by the more sensitive experiment. So what we really
care about is the situation with $g \approx 1$. We will set $g = 1$ in the following
derivation and calculation, and discuss $g \neq 1$ later.

To illustrate Eq.~(\ref{eq:main}) further, we
assume we know the background very well then simplify $f_0(\lambda_B) = \delta(\lambda_B -
m_B)$ ($\delta$ is the Dirac delta function and $m_B$ is the expected background). Then
Eq.~(\ref{eq:main}) can be rewritten as:
\begin{equation}
\label{eq:app2}
  f(\lambda_{S} | x_2, f_0(\lambda_{S},\lambda_{B2}))
=
  \frac{e^{-(2\lambda_{S}+m_{B1}+m_{B2})} (\lambda_{S}+m_{B1})^{x_1}
        (\lambda_{S}+m_{B2})^{x_2} }
       {\int e^{-(2\lambda_{S}+m_{B1}+m_{B2})} (\lambda_{S}+m_{B1})^{x_1}
        (\lambda_{S}+m_{B2})^{x_2} \mathrm{d}\lambda_{S}}
 \propto
  e^{-(2\lambda_{S}+m_{B1}+m_{B2})} (\lambda_{S}+m_{B1})^{x_1} (\lambda_{S}+m_{B2})^{x_2}
\end{equation}

\begin{multicols}{2}

From Eq.~(\ref{eq:app2}), it is easy to see in this simplified case
 that the posterior of the combined results of two measurements is just
proportional to the product of the posterior of each single one.
We can then infer that a prior distribution with a more pronounced
peak at $\lambda_S=0$ will produce more stringent posterior if
combined with the same secondary experiment. This means that a more
accurate measurement will play a dominant role in the combined
result. Furthermore, from Eq.~(\ref{eq:ori_main})
or~(\ref{eq:app2}) it is obvious that if we switch the sequence
of any two experiments, {\it i.e.} change the prior candidate for the other, the final result does not change. This
feature is natural and intuitive, and is advantageous compared with
the Serialization method~\cite{Yellin:2011}, whose results depend on
the sequence of the combined experiments.

\section{Numerical illustration}
Now let us consider a more practical situation, in which the
estimated backgrounds are assumed to satisfy a Gaussian distribution
whose mean value $m_B$ is same as that in the previous $\delta$
function while its standard deviation $\sigma_{B}$ depends on the
uncertainty of the estimation method. Eq.~(\ref{eq:main}) will be
transformed into a more complex function, so we composed a program,
named CULBA, based on C$++$ and ROOT's~\cite{ROOT} built-in
functions to implement the integration in Eq.~(\ref{eq:bay2}) and
later calculations as well as plotting. To simplify plots and discussion, we set $g=1$ during the numerical illustration without losing the general features of the results as discussed in Sec.~\ref{sec:com}. Here we choose the numerical
method instead of analytical expressions such as that used in
Ref.~\cite{Loparco:2011yr}, because the numerical method will make
our program more flexible to handle more types of priors in case the
exact formulae are missing or very complex. To illustrate the
functions of this program and its basic idea, we suppose there are
three independent measurements, I, II and III, and list
the data sets of the numbers of observed ($x$) and backgrounds ($\lambda_B$) in the signal region
respectively in Table~\ref{tab:single}. When the observed events $x$ is not significantly
larger than the expected backgrounds, just the upper limits of these
measurements are determined. Fig.~\ref{fig:upp1} shows the posterior
and their cumulative distributions with the uniform prior of the
three data sets; the upper limits are determined at $90\%$ credible
level (C.L.). As we mentioned before, we can use one result as
an input prior to calculate the posterior of the other one, as shown
with Eq.~(\ref{eq:main}), we refer to this as a ``transfer prior''.

Upper limit results of each single measurement and the combined ones obtained by our method are listed in
Tables~\ref{tab:single} and ~\ref{tab:combined}, respectively. From
these, two further intuitive features of the upper limit combination
based on the Bayesian method are deduced: 1) a more stringent final
result is expected when the results of two measurements are
combined, especially if these two measurements are at same precision
level; 2) if one measurement is much more precise than the other
, then the final combined result depends dominantly on the
more precise one.
For comparison
we use an ``experimentally practical'' method, which is widely applied in high energy physics analysis such as~\cite{Ablikim:2014fpb,Ablikim:2014atq}, to calculate upper limits and combine any two results. All the results based on the these methods
are listed in Tables~\ref{tab:single} and ~\ref{tab:combined} for comparison, and it turns out the ``experimental practical'' method provide
similar results to ours for both single and combined upper limits.
\end{multicols}

\begin{center}
\figcaption{\label{fig:upp1} PDF and cumulative
distribution functions of
  sets I, II and III, where the uniform prior is applied. The dotted, dashed and
  solid lines correspond to sets I, II and III respectively. The input values of these sets are
  shown in Table~\ref{tab:single}.}
\includegraphics[width=0.8\textwidth]{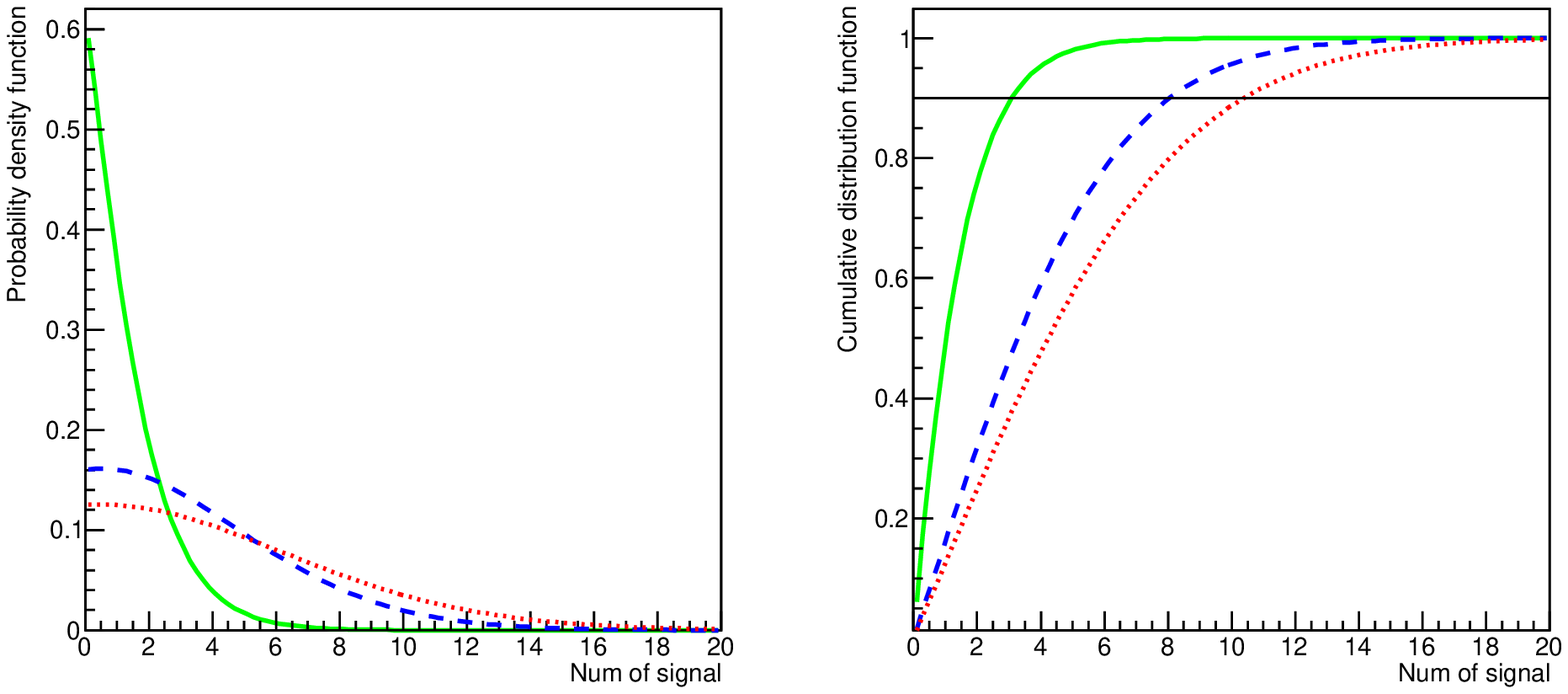}
\end{center}

\begin{multicols}{2}
\begin{center}
\tabcaption{ \label{tab:single} The three data sets and upper
limits (UL) at $90\%$ credible level based on Bayesian method with
 uniform prior (uni). For comparison, the ``experimentally practical'' method (exp) is also applied to
calculate the upper limits at $90\%$ credible level.}
\footnotesize
\begin{tabular*}{0.45\textwidth}{@{\extracolsep{\fill}} c|c|c|c}
\toprule
Mea.  &  I ($x/\lambda_B/\sigma_B$)   &      II ($x/\lambda_B/\sigma_B$)     &  III($x/\lambda_B/\sigma_B$)      \\
             &  $16/16/4$                    &       $9/9/3$                         &  $1/4/2$                         \\\hline
UL(uni)  &  $10.38$  &       $8.05$ & $3.09$ \\\hline
UL(exp)  &   $10.56$     &   $8.24$ &   $2.99$        \\
\bottomrule
\end{tabular*}
\end{center}

Further than this simple illustration, a systematic study with
different combinations of general conditions of experiments are
implemented by using CULBA. For two experiments I and II, we vary
their observations $x$ and expected backgrounds $\lambda_B$ from $0$
to $9$ with a step $1$ respectively, while the uncertainty of the
background is taken as $1$ when $\lambda_B=0$ or $\sqrt{\lambda_B}$
when $\lambda_B \geq 1$ for simplicity. Experiment I uses the uniform
prior, then its posterior PDF is used as the prior input to Experiment
II. The combined upper limits at $90\%$ C.L. for the
$(10\times10)^2=10000$ different combinations are calculated, and
they are shown in
Figs.~\ref{fig:combined_results1}-~\ref{fig:combined_results5}
classified by the observation $x^{II}$ of experiment II. In each
plot, y-axis is the upper limit at $90\%$ C.L. and the x-axis is the
experimental condition type, {\it i.e.} $10*x^I+\lambda^I_B$. The
upper limits for each single experiment I are also drawn in these
plots for comparison.

\begin{center}
\tabcaption{ \label{tab:combined} The three data sets and
combined upper limits (UL) at $90\%$ credible level based on the
Bayesian method with transfer prior (trans) are presented. For
comparison, the ``experimentally practical'' method (exp) is also applied to
calculate the upper limits at $90\%$ credible level.}
\footnotesize
\begin{tabular*}{0.45\textwidth}{@{\extracolsep{\fill}}c|ccc}
\toprule
Mea.  &  I ($x/\lambda_B/\sigma_B$)      &  II ($x/\lambda_B/\sigma_B$)    &  III($x/\lambda_B/\sigma_B$)    \\
             &  $16/16/4$                       &  $9/9/3$                        &  $1/4/2$                         \\\hline
Com.  &  I + II                          &  II + III                       &  III + I                         \\\hline
UL(tra)      &  $6.28$                          &  $2.77$                         & $2.88$  \\\hline
UL(exp)       &  $6.98$                          &  $2.89$                         & $2.96$  \\

\bottomrule
\end{tabular*}
\end{center}

\end{multicols}

\begin{center}
\figcaption{Upper limits at $90\%$ C.L. with $x^{II}=0$ and
$x^{II}=1$. In each plot, $10\times 10 \times 10 = 1000$ different
combined conditions are considered, where the observation $x^I$ and
expected background $\lambda^I_B$ of experiment I, and the
expected background $\lambda^{II}_B$ of experiment II are varied
from $0$ to $9$ respectively. The upper limit of a single experiment
is also provided for comparison. \label{fig:combined_results1}}
\includegraphics[width=0.48\textwidth]{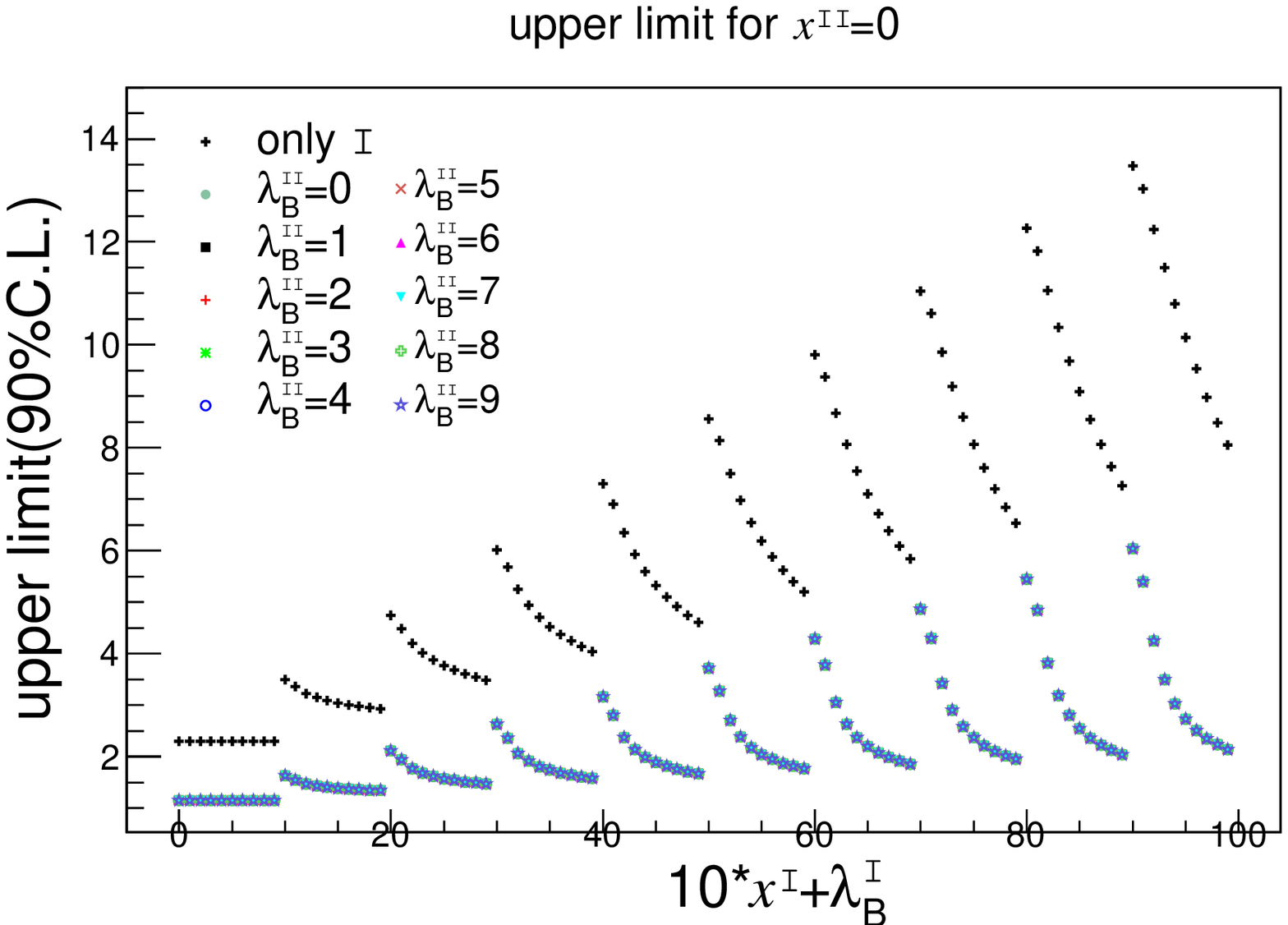}
\includegraphics[width=0.48\textwidth]{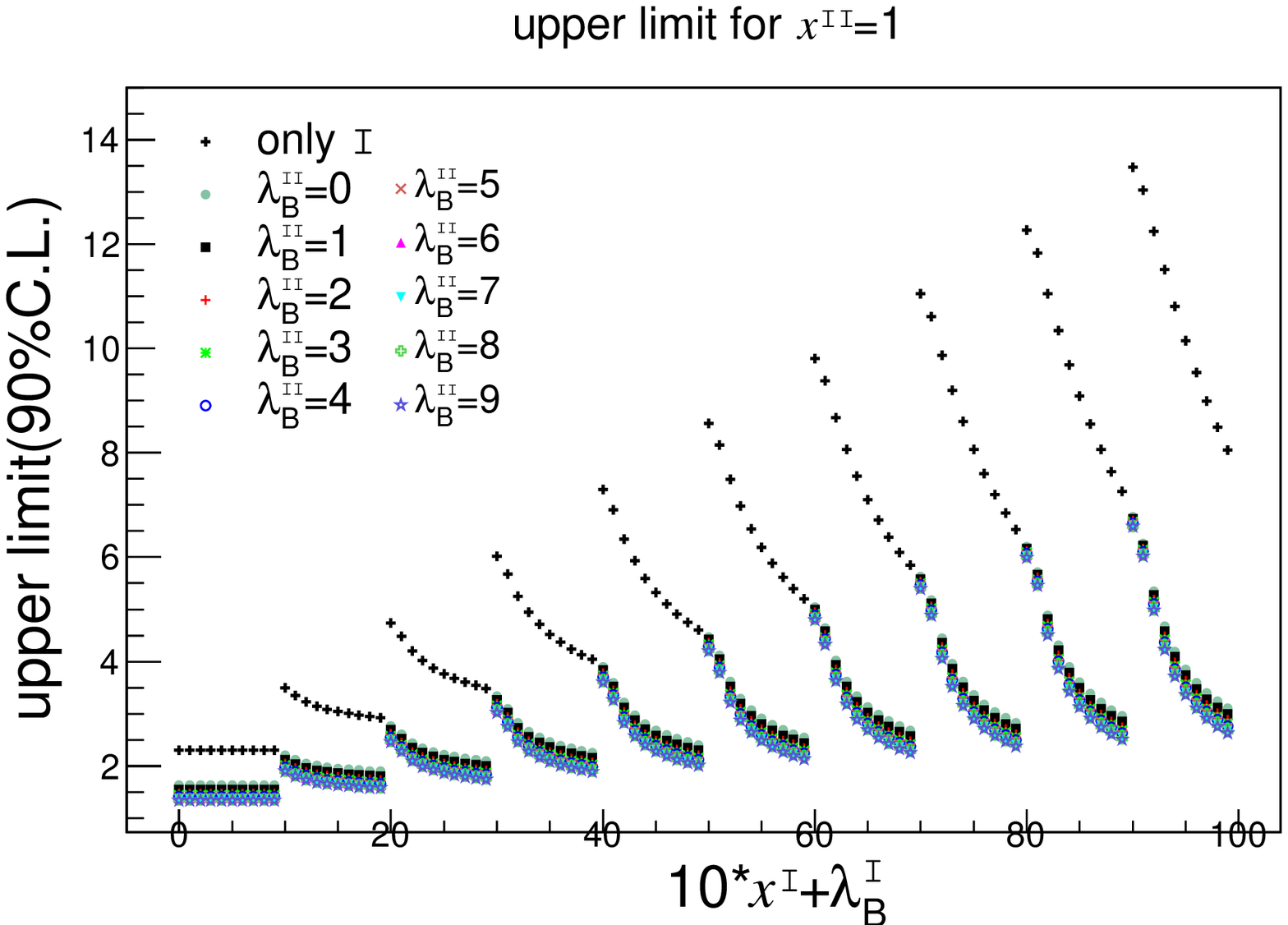}
\end{center}

\begin{center}
\figcaption{Upper limits at $90\%$ C.L. with $x^{II}=2$ and
$x^{II}=3$. }
\includegraphics[width=0.48\textwidth]{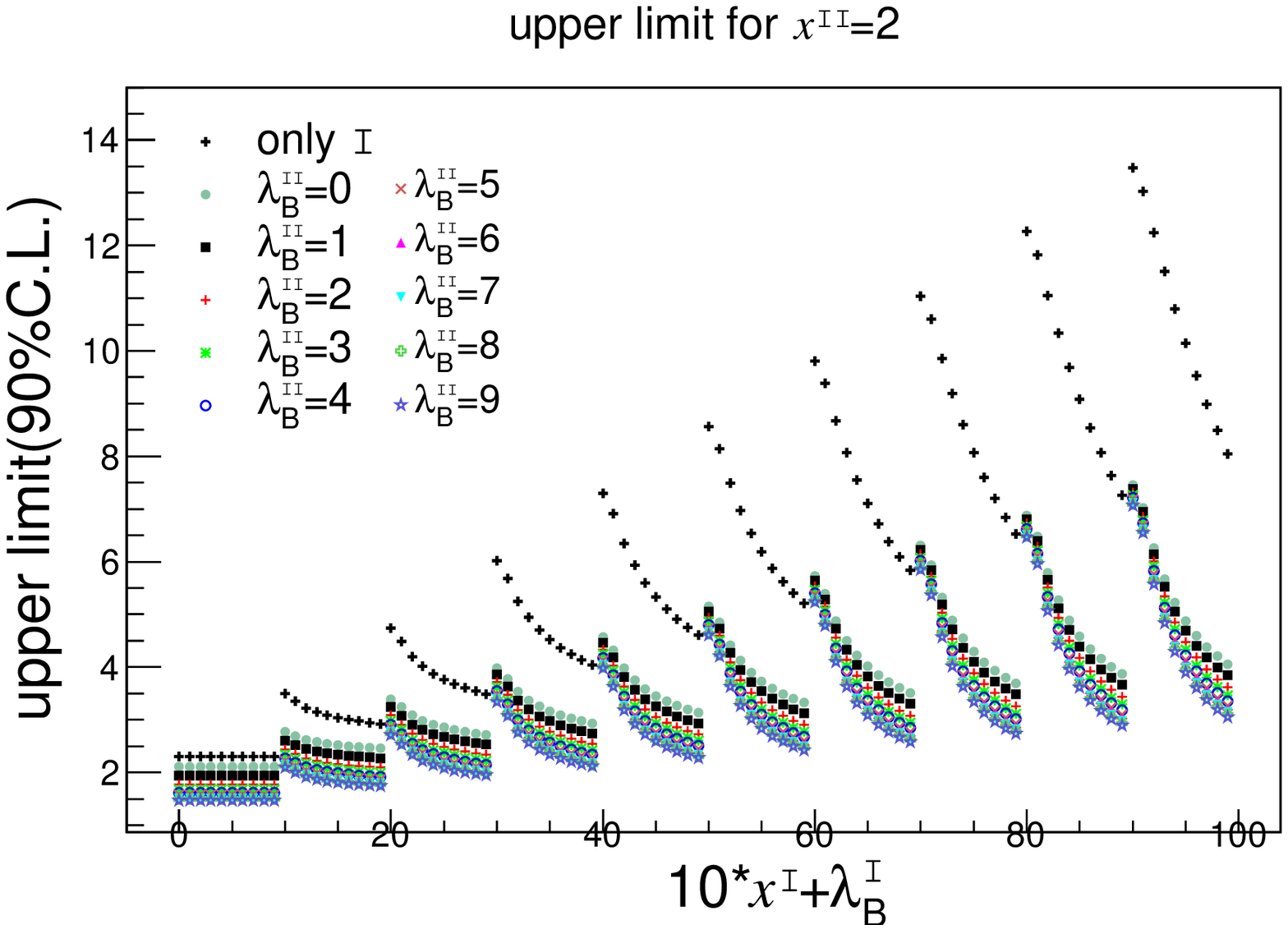}
\includegraphics[width=0.48\textwidth]{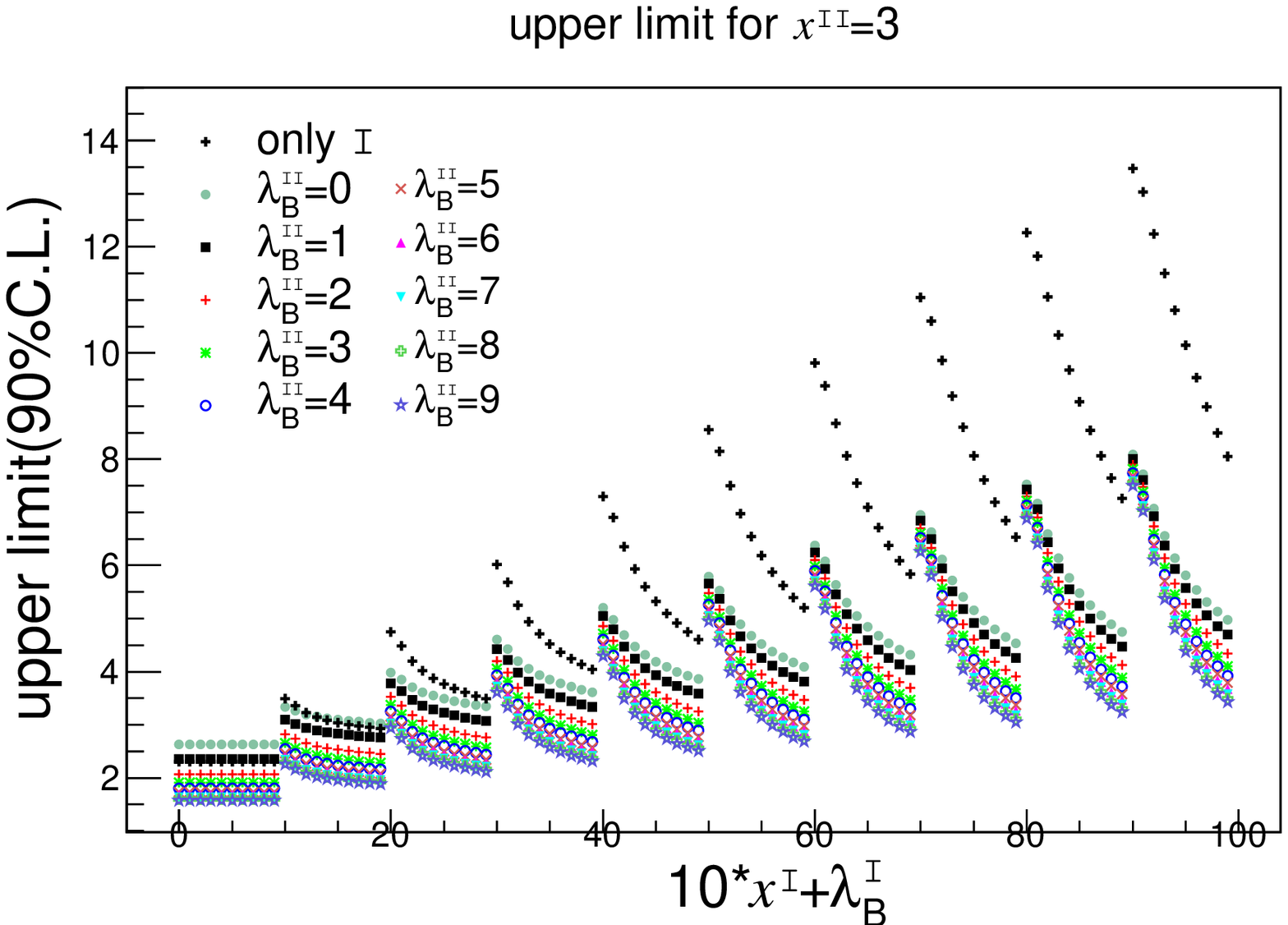}
\end{center}

\begin{center}
\figcaption{Upper limits at $90\%$ C.L. with $x^{II}=4$ and
$x^{II}=5$. }
\includegraphics[width=0.48\textwidth]{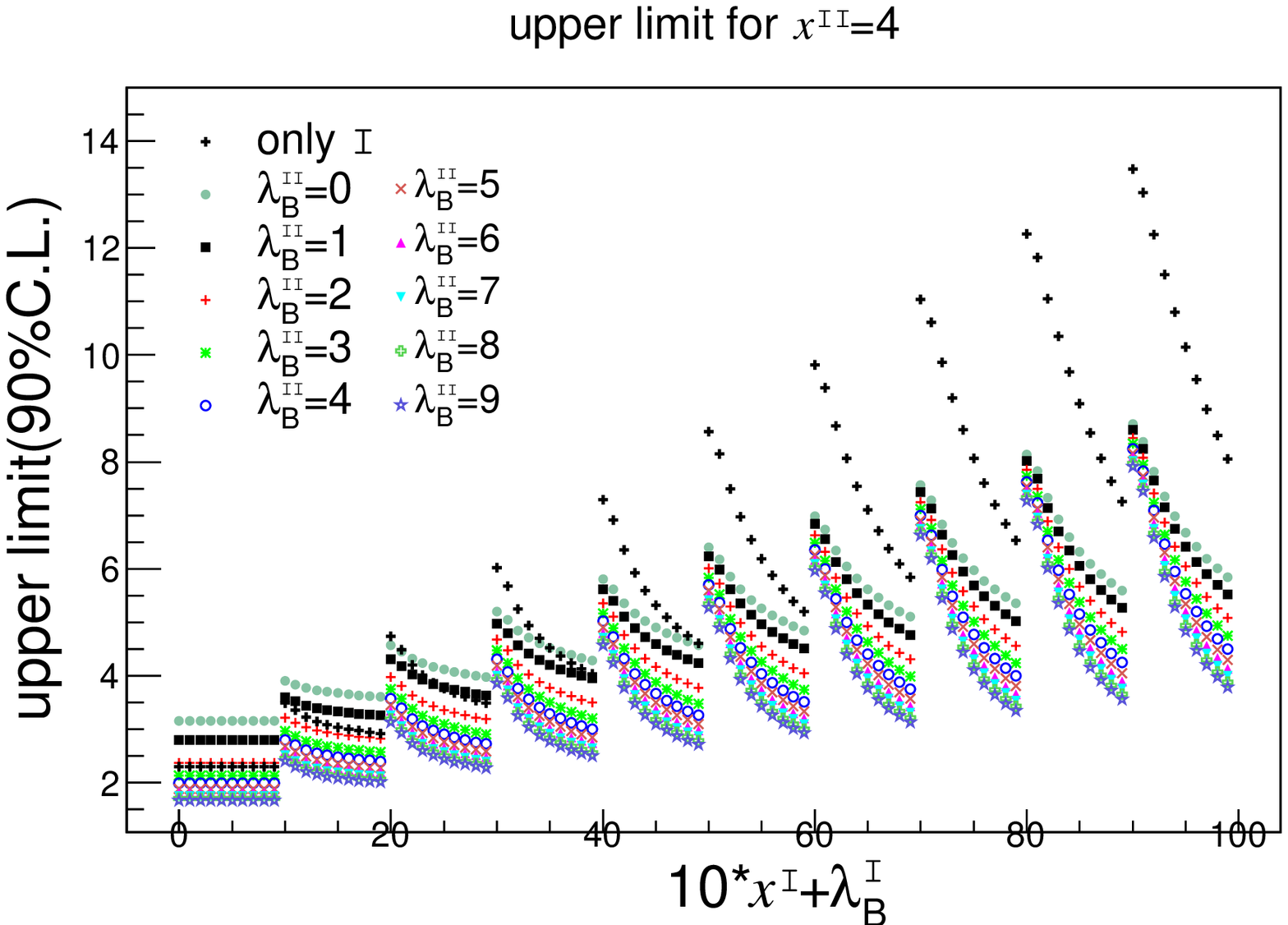}
\includegraphics[width=0.48\textwidth]{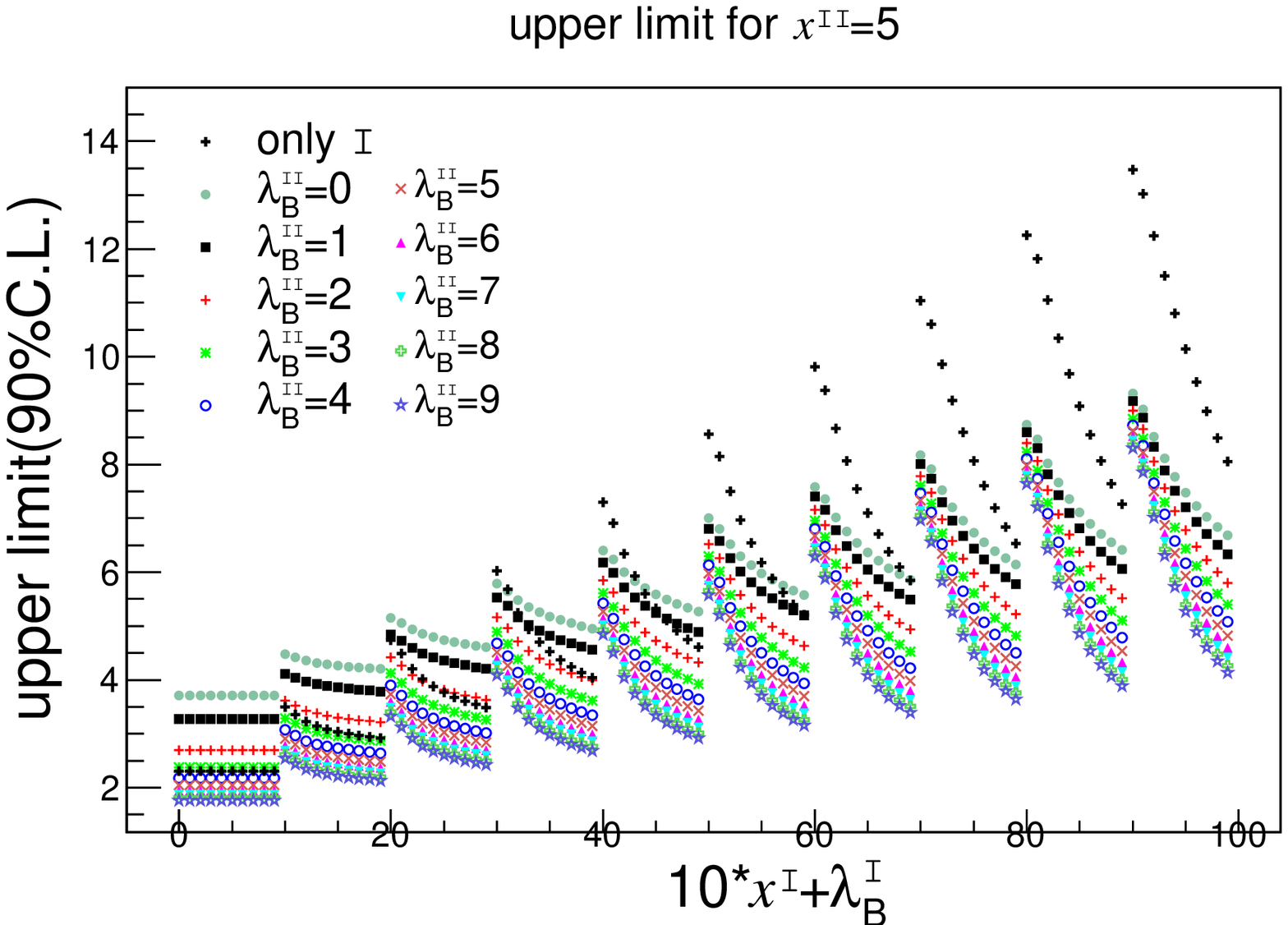}
\end{center}

\begin{center}
\figcaption{Upper limits at $90\%$ C.L. with $x^{II}=6$ and
$x^{II}=7$. }
\includegraphics[width=0.48\textwidth]{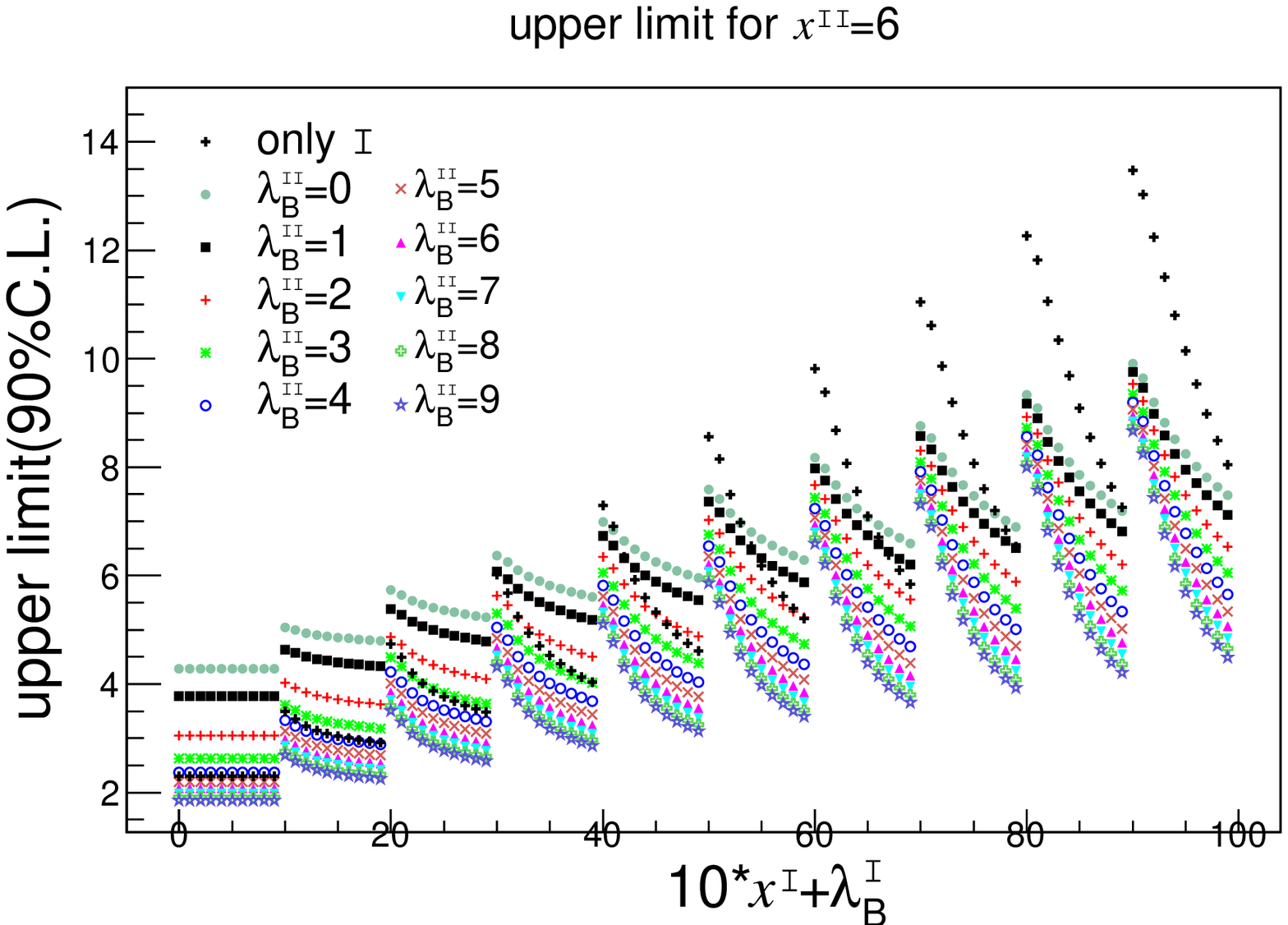}
\includegraphics[width=0.48\textwidth]{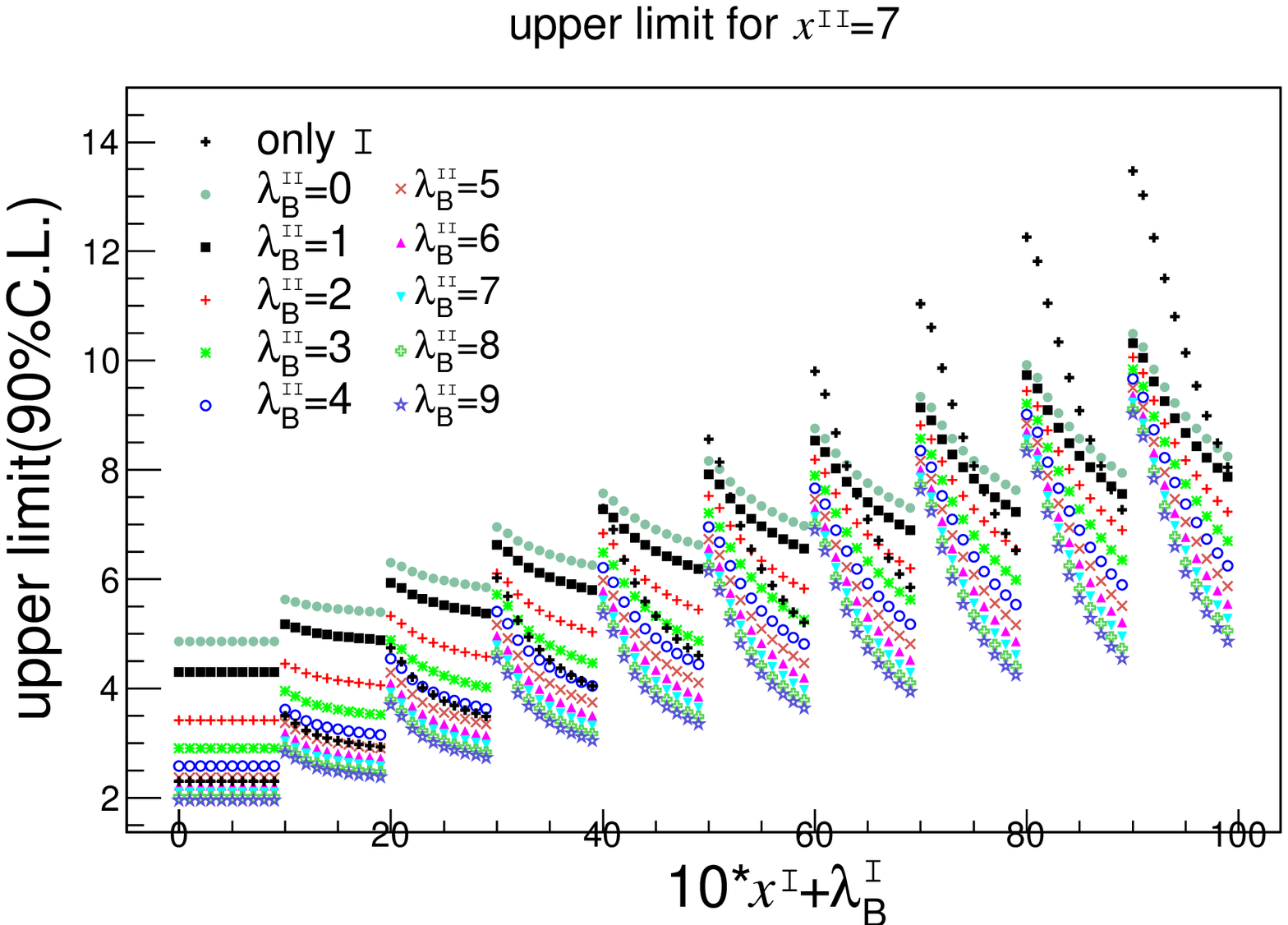}
\end{center}

\begin{center}
\figcaption{Upper limits at $90\%$ C.L. with $x^{II}=8$ and
$x^{II}=9$. }
\includegraphics[width=0.48\textwidth]{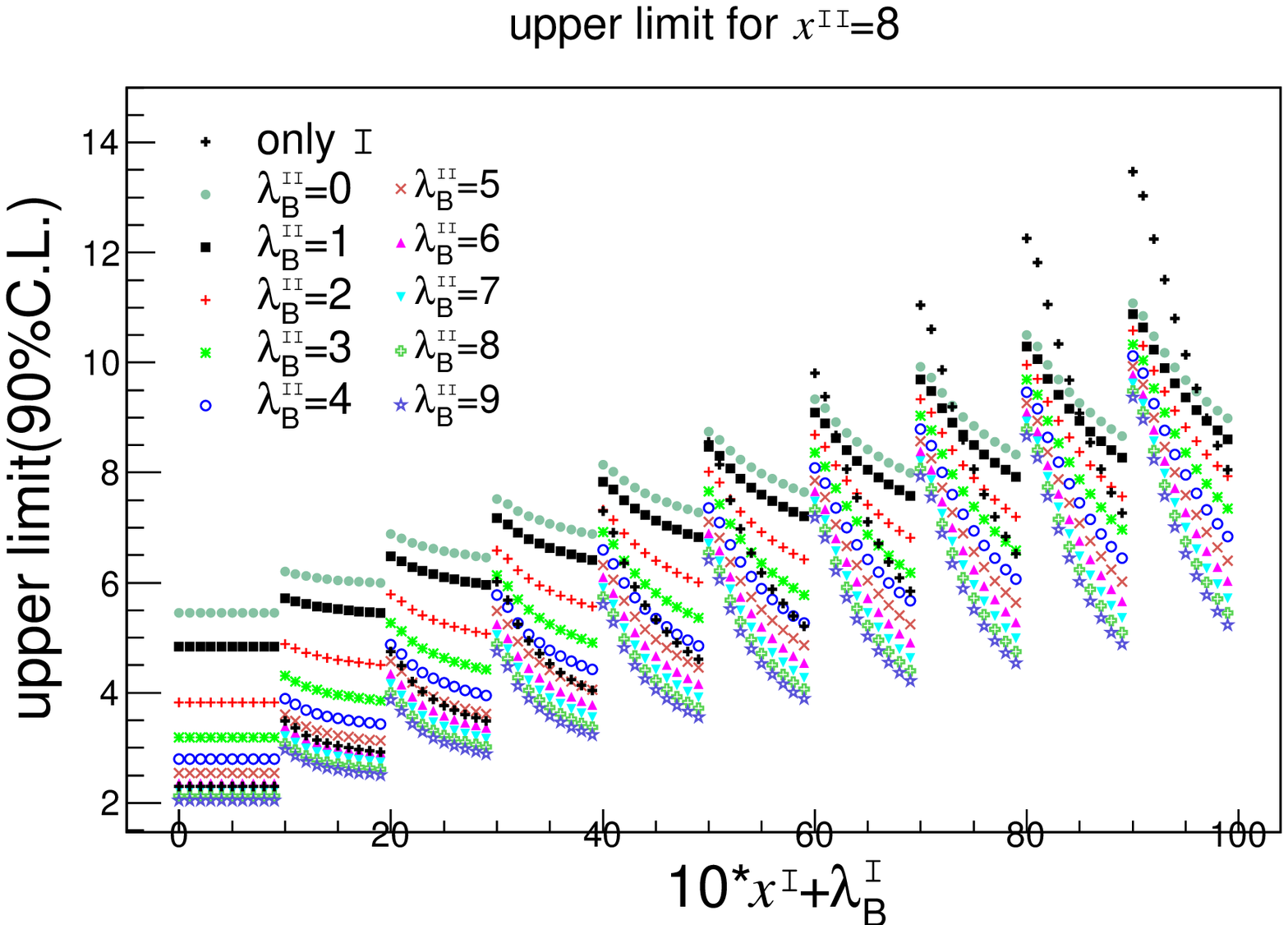}
\includegraphics[width=0.48\textwidth]{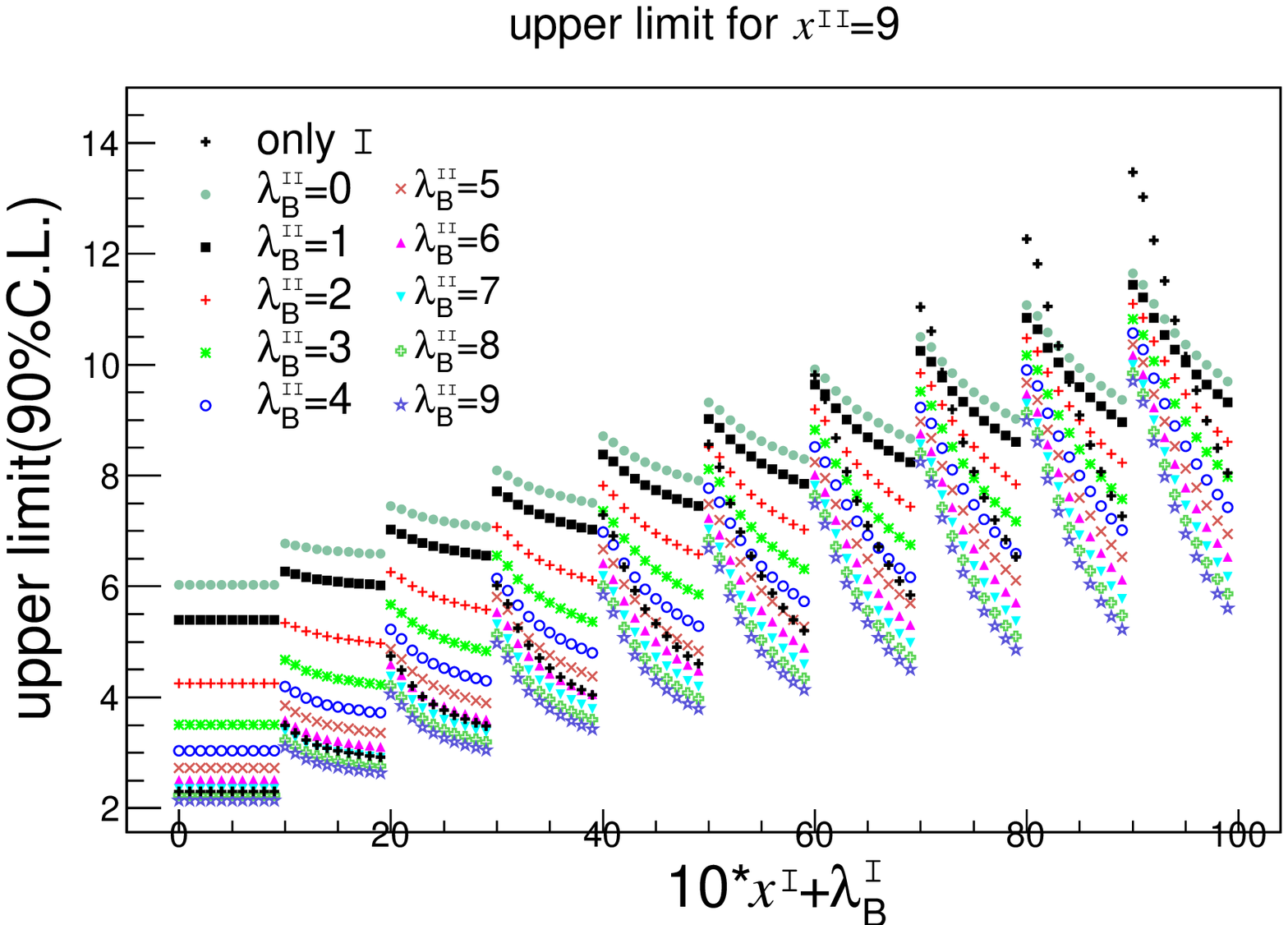}
\end{center}

\begin{center}
\figcaption{Comparison of the upper limits of single experiments
I (represented by red solid circles) and their combinations
  with experiment II (represented by grey stars) after grouping.
  The results are normalized to the average of each
  group. \label{fig:lsd}}
\includegraphics[width=0.58\textwidth]{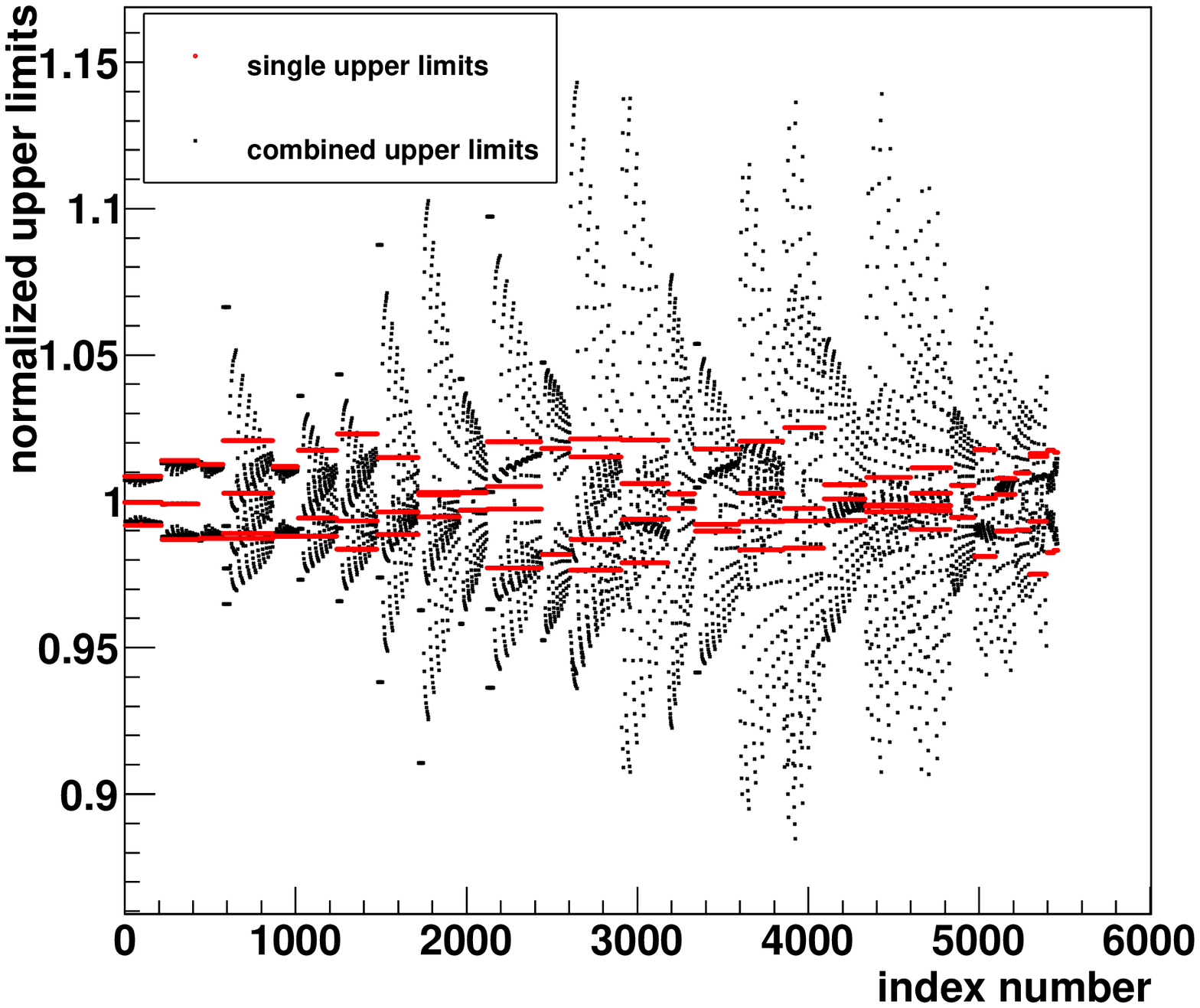}
\end{center}

\begin{multicols}{2}
\section{Discussion}
From the numerical results, all the four expected features mentioned in
Sec.~\ref{sec:com}, {\it i.e.}  ``switchable'', ``transitive'', ``improvable'', and
``dominant'', are observed. But we should notice that sometimes the combined upper limit
will be larger than the result from a single measurement. That happens only if the
observation is large and background is small, for example $x=5$ while $\lambda_B=0$. This
situation just indicates a significant signal is observed and an upper limit claim is not
proper anymore. A Bayesian method is available to deal with these conditions and is able to
provide combined mean values instead of upper limits, but that is out of the scope of this
paper so is not discussed here. We also notice not only the relative but the absolute numbers of signal and backgrounds are meaningful, because they will lead to different PDF shapes then different combination results. Another interesting topic is the dependence of likelihood
shapes, {\it i.e.}  when some individual experiments with different observations and
backgrounds give similar upper limits, will their combinations with another experiment
give similar results too? To study this, we divided the experiments I into different
groups with respect to their single upper limits by requiring the difference between any
two experiments to be less than $5\%$ inside each group. For instance, assuming there are seven
experimental conditions with upper limits of $1.00$, $1.02$, $2.01$, $2.05$, $4.23$,
$4.37$, and $4.44$ respectively, we divide them into three groups ($1.00$, $1.02$),
($2.01$, $2.05$), ($4.23$, $4.37$, $4.44$) by requiring the difference between any two
experiments in one group to be less than $5\%$.  Then within each group, we combined
experiments I with experiments II to obtain the combined upper limits. After that,
all the single and combined upper limits are normalized to the average of the group they
belong to respectively. The results are shown in Fig.~\ref{fig:lsd} for comparison, where the results from all groups are included. Fig.~\ref{fig:lsd}  shows that when the
difference of individual experiments is limited to $5\%$, mostly the difference of combined
results are within $10\%$ while the largest difference reaches $25\%$. That means in a
situation where only the upper limits are known but without the detailed information for
the signal and background, we still can use this method to get a reasonable combined upper
limit with a larger uncertainty. This conclusion is also suitable for the situation with
the normalization factor $g \neq 1$ but $g \approx 1$. Here, the upper limit of the first
experiment should be normalized to the second one by considering the experimental factors
such as luminosities and efficiencies, then an approximate combined result can be
obtained with this method.

All the relevant numerical results can
be used as a reference to combine two experimental results appropriately. They are saved
to a plain text file and upload to the arxiv server as ``other formats''~\cite{zhuk:2013}.

Notice that we only discussed combining two measurements in this article, but with the Bayesian
method it is simple and easy to expand to any number of measurements with a combination
chain, where each result in the previous step will be used as an input prior for the next
step.

\vspace{10mm} \centerline{\rule{80mm}{0.1pt}} 
\end{multicols}

\clearpage

\end{document}